\documentclass[journal=jacsat,manuscript=article]{achemso}
\usepackage[version=3]{mhchem} 
\usepackage{siunitx}
\usepackage[dvipsnames]{xcolor}
\usepackage{amssymb}
\usepackage{pifont}
\usepackage{wrapfig}
\usepackage{hyperref}

\usepackage{amsmath} 

\usepackage{tabularx,colortbl}
\usepackage{capt-of}
\usepackage{adjustbox}
\usepackage{subcaption}

\author{Garima Chib}
\affiliation[meche]
{Department of Chemical Engineering, Carnegie Mellon University, 15213, USA}

\author{Parisa Mollaei}
\affiliation[meche]
{Department of Mechanical Engineering, Carnegie Mellon University, 15213, USA}

\author{Amir Barati Farimani}
\email{barati@cmu.edu}
\affiliation[meche]
{Department of Mechanical Engineering, Carnegie Mellon University, 15213, USA}
\alsoaffiliation[biomed]
{Department of Biomedical Engineering, Carnegie Mellon University, 15213, USA}
\alsoaffiliation[mld]
{Machine Learning Department, Carnegie Mellon University, 15213, USA}

\title[An \textsf{achemso} demo]
{Characterizing the Conformational States of G Protein Coupled Receptors Generated with AlphaFold}

\abbreviations{IR,NMR,UV}

\begin{document}
\begin{abstract}

G-Protein Coupled Receptors (GPCRs) are integral to numerous physiological processes and are the target of approximately one-third of FDA-approved therapeutics. Despite their significance, only a limited subset of GPCRs has been successfully targeted, primarily due to challenges in accurately modeling their structures. AlphaFold, a state-of-the-art deep learning model, has demonstrated remarkable capability in predicting protein structures with high accuracy. This study conducts an evaluation of AlphaFold’s performance in predicting GPCR structures and their conformational states by comparing its predictions to experimentally determined structures using metrics such as average deformation between alpha carbon atoms and the Helix 3 - Helix 6 (H3-H6) distance. Our analysis reveals that both AlphaFold 2 (AF2) and AlphaFold 3 (AF3) produce more accurate predictions for GPCRs in inactive conformations, with lower activity levels correlating with smaller deformations. Conversely, higher activity levels are associated with increased variability in AlphaFold performance due to difficulties with accurately predicting conformational changes upon GPCR activation and ligand binding. Additionally, AlphaFold’s performance varies across different GPCR classes, influenced by the availability and quality of training data as well as the structural complexity and diversity of the receptors. These findings demonstrate the potential of AlphaFold in advancing drug discovery efforts, while also highlighting the necessity for continued refinement to enhance predictive accuracy for active conformations.

\end{abstract}

\clearpage

\section{Introduction}

G-Protein Coupled Receptors (GPCRs) form a large family of cell-surface receptors that play a vital role in multiple signaling pathways\cite{FOSTER2019895, JI199817299, LIEBMANN2001777, blesen1995receptor, GUTKIND19981839, BETKE2012304}. GPCRs are involved in vision, taste, and immune system regulation, among other physiological functions\cite{Rosenbaum2009-fn}. As a result of their role in mediating most cellular responses to hormones, approximately one-third of U.S. Food and Drug Administration (FDA) approved drugs target GPCRs \cite{Sriram251, Lagerstrom2008-ri, HAUSER201841}. Despite their critical role in mediating cellular responses to hormones and their prominence as drug targets, advancing GPCR research remains challenging due to the complexity of determining their structures and conformational dynamics\cite{CONGREVE202081, Dill2008}. Experimental efforts to determine protein structures such as X-ray crystallography, nuclear magnetic resonance (NMR), and cryo-electron microscopy suffer from long timelines, trial and error processes, high costs, and technical limitations, including difficulties in crystallizing membrane proteins, and low-resolution constraints. \cite{Thompson2020, Bai_McMullan_Scheres_2015, Wthrich2001, Jaskolski_Dauter_Wlodawer_2014a} To overcome these challenges, computational methods capable of accurately predicting protein structures from its amino acid sequence offer a powerful alternative, significantly accelerating drug discovery efforts. \cite{Radivojac2013} Traditional computational approaches for protein structure prediction, such as protein-protein interaction networks, \cite{Deng2003, Letovsky2003, Vazquez2003} inferred evolutionary correlations, \cite{Pellegrini1999, Marcotte1999, Enault2005} and physics-based simulations\cite{Brini2020, Sippl1990} have contributed valuable insights. However, these methods fall short of experimental accuracy, which limits their utility for a broad range of biological applications. \cite{Jumper2021}

In recent years, there have been efforts to incorporate machine learning in protein structure prediction. \cite{Senior2020, Wang2017, Zheng2019, LeCun2015, Goodfellow-et-al-2016, Radford2019LanguageMA, OFER20211750} AlphaFold is a state-of-the-art deep learning system that can generate highly accurate predictions of the 3D structure of proteins. \cite{Jumper2021, Nussinov2022, jumper2021highly}. It is the first computational approach to protein structure prediction that has achieved near experimental accuracy in the majority of cases. \cite{Jumper2021, Kryshtafovych2021} AlphaFold incorporates biological and geometric information about protein structure into its deep learning algorithm. It constructs a multi-sequence alignment (MSA) that identifies sequences similar to the input sequence. \cite{Jumper2021} This is then passed through an Evoformer, which is the building block of the transformer-based model architecture. \cite{Jumper2021, Huang_2019_ICCV} This allows the model to learn spatial and evolutionary relationships between amino acids. With AlphaFold, it is possible to generate the 3D conformations of GPCRs even when experimental results are unavailable. Given that AlphaFold can potentially predict the structure of any GPCR with high accuracy, it holds significant promise for advancing drug development targeting a diverse array of GPCRs. \cite{Nussinov2022} Consequently, a comprehensive evaluation of AlphaFold's predictions is crucial to assess its effectiveness in modeling GPCR structures.

\begin{figure}[t!]
    \centering
    \includegraphics[width=0.8\linewidth]{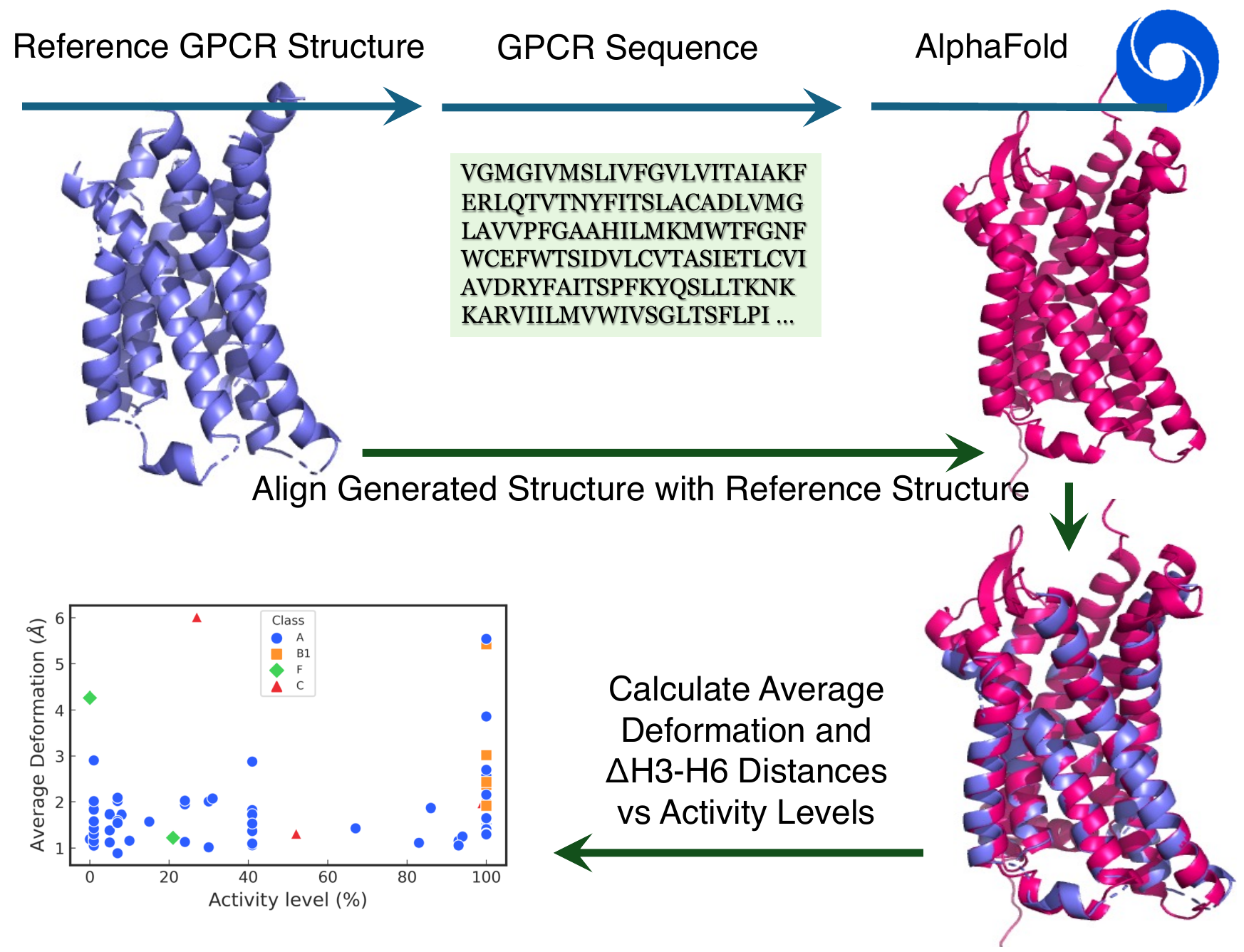}
    \caption{Framework for assessing the predictive accuracy of AlphaFold in relation to GPCR activity levels. It consists of extracting the GPCR sequence from a reference structure, using it as input to AlphaFold, aligning the predicted and reference structures, and computing structural metrics such as average deformation and $\Delta$H3-H6 distances.}
    \label{fig:framework_figure}
\end{figure}

When a ligand binds to the extracellular part of a GPCR, a series of conformational changes are triggered in the receptor as it becomes activated.\cite{Zhang2015, Hilger2018, Kohlhoff2013} These changes are then relayed across the protein to the intracellular region, causing G-protein coupling. \cite{Hoffmann2008} Impairments to these signaling pathways in GPCRs are associated with damage to the central nervous system, cardiovascular diseases, and cancer. \cite{Sriram251, CONGREVE202081} Conformational changes in GPCRs represent a critical area of research in drug design and development. \cite{Hassan2004, Lee2001, Deupi2010, Trzaskowski2012} In this study, we evaluate AlphaFold's ability to predict GPCR conformations linked to activity levels ranging from 0 to 100 percent. Assessing its accuracy in predicting these conformational changes could underscore AlphaFold's potential in developing more effective therapeutics.
To assess AlphaFold's predictive accuracy, we define the deformation between ground truth and generated structures as the average distance between alpha carbon atoms in both structures. We also measured contact distances between two residues in the intracellular regions of the third and sixth transmembrane helices (TM3 and TM6) as this distance is highly correlated with the activity level of GPCRs.\cite{Rasmussen2011, mollaei2023activity}. We also conduct a comparative analysis between AlphaFold 2 (AF2) and AlphaFold 3 (AF3) predictions. In addition, we examine deformation trends across different GPCR classes to identify any patterns in AlphaFold's performance. GPCRs are categorized into several classes based on their structure and function. \cite{Odoemelam2020} 

Class A (Rhodopsin) GPCRs, constituting about 80\% of human GPCRs, share specific sequence motifs such as the DRY motif and NSxxNPxxY motif, and have seven transmembrane helices forming an elliptical, cylindrical shape. \cite{Schith2005, Rosenbaum2009-fn, BGacasan2017} Despite the diversity in ligand preferences and primary structures, they have similar N-terminal regions and conserved motifs in the transmembrane domains. \cite{BGacasan2017, Palczewski2006} Class B (Secretin) GPCRs possess an extracellular hormone-binding site and a conserved extracellular N-terminal domain (ECD) essential for their function, using a `two-domain' binding mechanism. \cite{Lagerstrom2008-ri, Wheatley2012, Odoemelam2020, deGraaf2017} Class C (Glutamate) GPCRs have a large extracellular domain with orthosteric sites and form dimers, characterized by a Venus flytrap (VFT) module.\cite{Lagerstrom2008-ri, Fredriksson2003, Chun2012} Class F (Frizzled) receptors, initially deemed unconventional, play critical roles in tissue development and physiological balance, forming complex assemblies called signalosomes. \cite{Schulte2019, Schulte2018, DeBruine2017} These structural and functional distinctions among GPCR classes are essential for evaluating AlphaFold's predictive accuracy across diverse receptor types.

\section{Methods}
Figure \ref{fig:framework_figure} outlines the overall framework for evaluating AlphaFold’s predictions across GPCR activity levels. The process involves extracting the GPCR sequence from a reference structure, generating a predicted structure using AlphaFold, and quantifying structural deviations through average deformation and $\Delta$H3-H6 distances.

\subsection{Dataset}
The ground truth GPCR structures consist of 75 GPCRs with four unique receptor classes obtained from the Protein Data Bank (PDB) database \cite{Berman2003}. The dataset contains 63 receptors in Class A, six receptors in Class B1, four receptors in Class C, and two receptors in Class F. For our analysis, we truncate these structures to retain only the seven transmembrane regions of the GPCRs as these domains are integral to the activation process. The activity level associated with each GPCR structure is obtained from the GPCRdb database. \cite{Isberg2013} This database consists of experimental information about GPCRs, including ligand binding, sequence alignments, and mutation analysis. The AF2 generated structures are obtained from the AlphaFold Protein Structure Database. The sequence overlap between the reference and generated sequences exceeds 96\% for all samples (see Section 1 in Supporting Information). Additionally, AF3 generated structures were obtained from the AF3 server. \cite{Abramson2024}

\subsection{Estimation of AlphaFold's prediction accuracy}
To estimate AlphaFold’s predictive accuracy, we align the ground truth and predicted structures using PyMOL, a molecular visualization software. \cite{PyMOL, delano2002} This process involves sequence alignment and spatial superposition. To calculate deformation, we utilize the Biopython package in Python, which offers a variety of tools for computational biology. The workflow includes the following steps: 1. Parsing the PDB files for both ground truth (reference) and predicted structures to iterate through the entire amino acid sequence. 2. Extracting the seven transmembrane (TM) regions from the reference structures. 3. Identifying the corresponding TM regions in the predicted structures that consist of the same sequence of residues. 4. Calculating the Euclidean distance between alpha carbon atoms of matching residues in the reference and predicted structures (see Section 3 in Supporting Information). The average distance between these alpha carbons is used as a metric to evaluate AlphaFold’s prediction accuracy.

Additionally, we use the H3-H6 distance\cite{mollaei2023activity, mollaei2023unveiling} as a metric to evaluate the error in AlphaFold-generated GPCR structures. To determine this distance, we identify the third-to-last residue in the third and the sixth TM region of each GPCR, calculating the Euclidean distance between their alpha carbons. To achieve this, we use the MDTraj Python library \cite{mcgibbon2015mdtraj}, which offers tools for analyzing molecular dynamics trajectories. We then calculate the absolute difference between the H3-H6 distance in the reference and predicted structures to assess the accuracy of AlphaFold predictions (see Figure S1 in Supporting Information).

To determine differences in AlphaFold performance across various GPCR classes, we refer to GPCR-BERT\cite{Kim2024}, a protein language model for interpreting GPCR sequences and extracting higher-order interactions in GPCRs. The BERT model incorporates a classification [CLS] token, which is used to categorize the sequence into certain classes. \cite{BERT} This encompasses all important information about the sequence.\cite{Schwaller2019, 2012.06051} By extracting the [CLS] token embeddings from the final hidden state of the GPCR-BERT model, we can perform a t-distributed stochastic neighbor embedding (t-SNE) analysis.\cite{JMLR:v9:vandermaaten08a} This is a dimensionality reduction technique that is used for visualizing high-dimensional data. It computes pairwise similarities between data points in a high-dimensional space and converts them to joint probability distributions. It then minimizes the Kullback-Leibler divergence between the probability distributions in the high-dimensional and low-dimensional space. This attracts similar data points towards each other and pushes dissimilar points further apart, allowing for the formation of clusters. Through this analysis, we can uncover variations in AlphaFold performance for different receptor classes. 

\section{Results and Discussion}

\begin{figure}[t!]
    \centering
    \includegraphics[width=\linewidth]{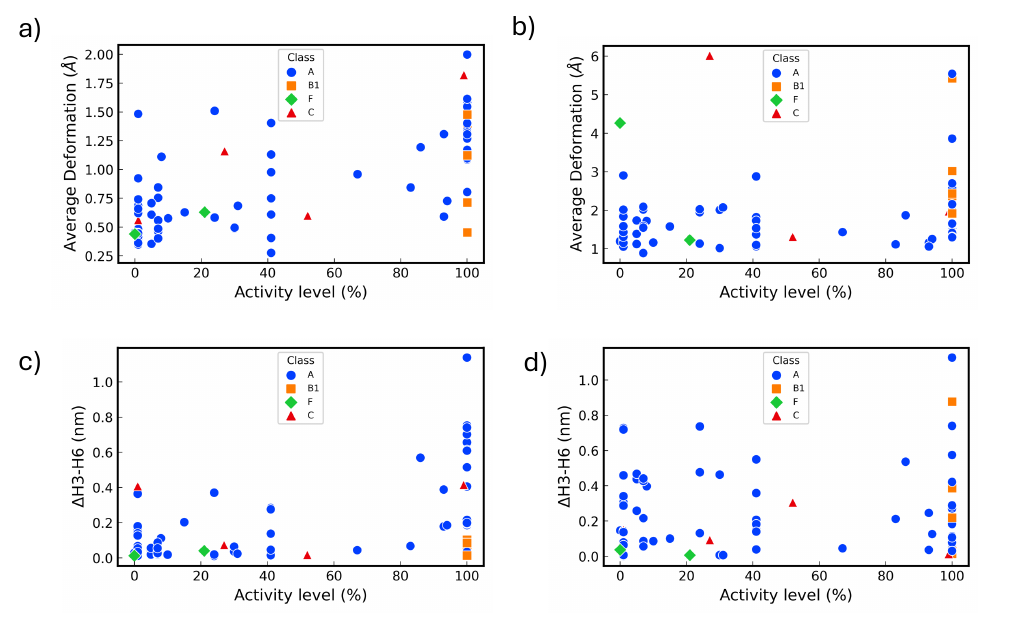}
    \caption{Analysis of average deformation and differences in H3-H6 distances in relation to activity levels. Figures (a) and (b) illustrate the average deformation results for AF2 and AF3, respectively. Figures (c) and (d) present the differences in H3-H6 distances for AF2 and AF3, respectively.}
    \label{fig:deform_h3_h6_results}
\end{figure}

We measured the average deformation of GPCR structures predicted by AF2 and AF3 and analyzed these results concerning activity levels as shown in Figure \ref{fig:deform_h3_h6_results}a and b. For AF2, the average deformation ranges from 0.25Å to 2Å. Our observations indicate that AF2 generates more accurate conformations for GPCRs with lower activity levels, while higher activity levels correspond to greater fluctuations in average deformation. Similarly, AF3 produces more accurate structures for inactive GPCR conformations. However, the range of average deformation values for structures predicted by AF3 is broader, extending from approximately 0.8Å to 6Å, indicating greater variability in performance compared to AF2. This suggests that AF2 more consistently predicts GPCR conformations with lower structural deformations. 

We also analyzed the differences in H3-H6 distances predicted by AF2 and AF3 concerning activity levels as depicted in Figure \ref{fig:deform_h3_h6_results}c and d. A consistent trend was observed in both models: the difference in H3-H6 distances is smaller for inactive conformations compared to active ones. The range of H3-H6 distance values generated by AF2 and AF3 is comparable. However, the $\Delta$H3-H6 values predicted by AF2 are consistently lower than those from AF3 for inactive GPCR conformations. This suggests that while AF3 may overestimate structural changes in inactive states, AF2 provides more accurate predictions for these conformations.

\begin{figure}[t!]
    \centering
    \includegraphics[width=\linewidth]{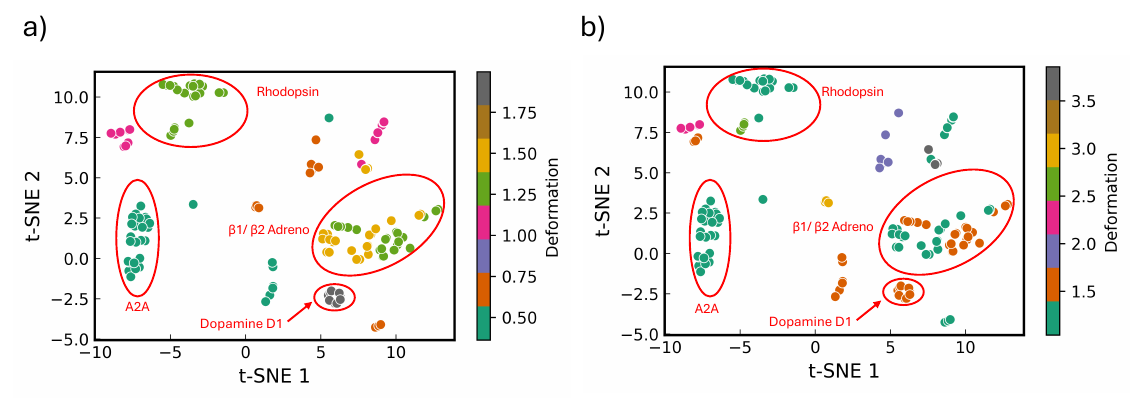}
    \caption{t-SNE visualization of AlphaFold performance across various GPCR classes using embeddings from GPCR-BERT. Clusters represent distinct GPCR classes, with colors denoting specific ranges of average deformation. Figure (a) illustrates the results for AF2, while Figure (b) presents the results for AF3.}
    \label{fig:tsne_results}
\end{figure}

We performed a comparative analysis of AlphaFold performance across different receptor classes by applying t-SNE to embeddings from the final hidden state generated by GPCR-BERT.\cite{Kim2024} This is demonstrated in Figure \ref{fig:tsne_results}. The resulting clusters correspond to distinct GPCR classes, with colors indicating specific ranges of deformation values. Our findings reveal that the adenosine A2A receptor exhibited the lowest deformation values for both AF2 and AF3. In contrast, the dopamine D1 receptor showed the highest deformation with AF2, while the serotonin 2A receptor had the highest deformation with AF3. It should be noted that the dataset used for the t-SNE analysis with GPCR-BERT is significantly smaller than the dataset used for our study (see Section 4 in Supporting Information). Consequently, the t-SNE plots do not encompass all the receptor classes included in our dataset.

AlphaFold demonstrates lower accuracy in predicting active GPCR conformations primarily due to the intrinsic dynamic nature of these states and biases in its training data. Active GPCRs undergo significant conformational changes upon ligand binding and intracellular coupling, resulting in a heterogeneous ensemble of structures rather than a single well-defined state. \cite{Hilger2018} This structural flexibility poses a challenge for AlphaFold, which is optimized for predicting static conformations. Furthermore, the scarcity of experimentally resolved active GPCR structures in databases such as the Protein Data Bank (PDB) contributes to prediction inaccuracies. \cite{CONGREVE202081} Inactive states are more frequently captured through techniques like X-ray crystallography and cryo-electron microscopy due to their higher stability, leading to an overrepresentation of these conformations in AlphaFold’s training data. \cite{Rasmussen2011} Furthermore, the structural differences and complexities between different GPCR classes can make it more challenging for AlphaFold to predict their structures accurately.

\section{Conclusion}

In this study, we highlight the efficacy of AlphaFold in predicting structures of GPCRs, a crucial class of proteins involved in numerous physiological processes. By analyzing the average deformation between ground truth and AlphaFold-generated structures, along with the H3-H6 distance, we assessed AlphaFold’s accuracy in predicting GPCR conformational states. Our results indicate that AF2 and AF3 generally produce more accurate structures for GPCRs with lower activity levels corresponding to smaller deformations. Higher activity levels are associated with greater structural fluctuations, highlighting the challenge of predicting active conformations accurately. AF3 exhibited a broader range of deformation values across all activity levels, indicating greater structural variability in its predictions. In contrast, AF2 demonstrated higher accuracy in predicting low-deformation structures and showed smaller deviations in H3-H6 distances for inactive states, suggesting more precise and reliable structural predictions. Furthermore, through t-SNE analysis of GPCR-BERT embeddings, we observed distinct clusters corresponding to different GPCR classes. The adenosine A2A receptor exhibited the lowest deformation values, while the dopamine D1 and serotonin 2A receptors showed higher deformation values. While AlphaFold demonstrates significant potential in accurately predicting GPCR structures, its performance varies depending on the receptor’s activity level, the availability of training data, and structural complexity. Continued refinement of these models can further improve their usefulness in drug discovery and development, especially for targeting the wide range of GPCRs that are potential drug targets.

\section{Data and software availability}
The necessary information containing the codes and data used in this study is available here: \url{https://github.com/garimachib01/GPCR_AlphaFold}

\section{Supporting Information}
The Supporting Information includes additional details on data preparation and calculation of deformation metrics.

\begin{acknowledgement}

This work is supported by the Center for Machine Learning in Health (CMLH) at Carnegie Mellon University and a start-up fund from the Mechanical Engineering Department at CMU. 

\end{acknowledgement}

\bibliography{reference}
\end{document}


\maketitle

\section{1: GPCRs Dataset Preparation}

The dataset contains 75 GPCRs along with their activity levels. We aligned the reference structures obtained from the RCSB PDB server with their AlphaFold generated counterparts using the align tool in PyMOL. \cite{PyMOL, delano2002} This allows us to calculate the average distance between corresponding alpha carbon atoms in both structures. 

We also calculate the ratio of overlapping sequences between the AlphaFold 2 generated structures and the reference structures. This is defined as the following:

\begin{equation} \label{eq:overlap_ratio}
    \text{Overlap Ratio} = \frac{\text{Length of Overlapping Sequence}}{\text{Length of Reference Sequence}}
\end{equation}

We calculate this using the pairwise alignment tool in the Biopython package. A concise overview of the process is given as follows:

\begin{enumerate}
  \item Extract sequences from both the sample and reference protein structures.
  \item  Perform global sequence alignment using the Bio.Align module
  \item Compare aligned residues and construct an overlapping sequence by retaining matching residues
  \item Calculate Overlap Ratio using Eq.\eqref{eq:overlap_ratio} for all GPCRs
\end{enumerate}

For structures generated by AlphaFold 3, the generated sequences are identical to the reference sequences.

\section{2: Visualization of H3-H6 Distances}

\begin{figure}[H]
    \centering
    \includegraphics[width = 16cm]{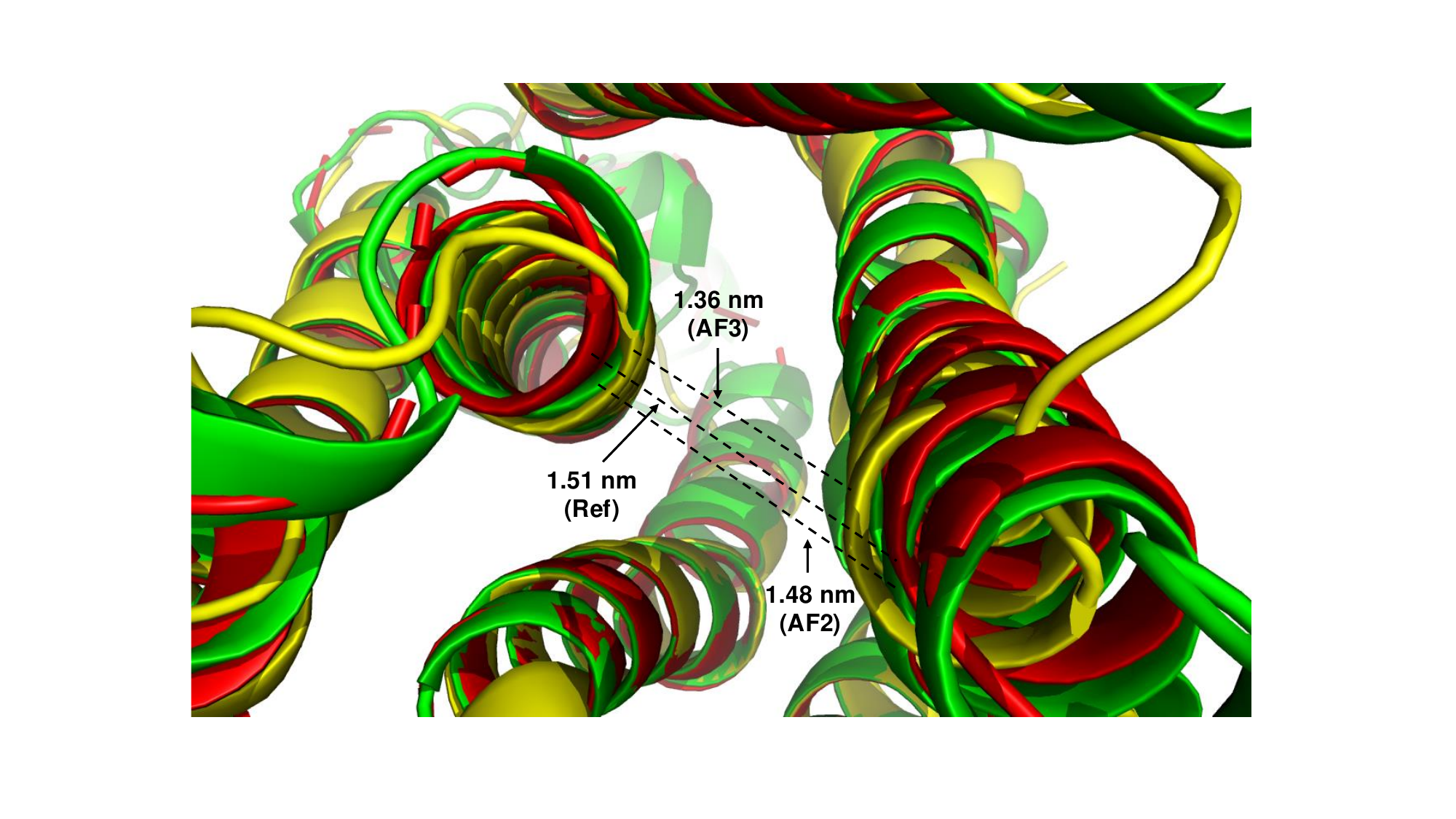}
    \caption{Demonstration of H3-H6 distances shown by dashed lines for the Adenosine A2A receptor. The ground truth structure is illustrated in red, AF2 generated structure in green, and AF3 generated structure in yellow.}
    \label{fig:h3_h6}
\end{figure}

Figure \ref{fig:h3_h6} shows a comparison of H3-H6 distances between the ground truth structure and predictions by AlphaFold 2 and AlphaFold 3 for the Adenosine A2A receptor. 

\section{3: Calculation of Average Deformation}

Since some of the AlphaFold 2 generated sequences do not exactly match the reference sequences, we face some additional challenges in calculating the deformation. A detailed description of the process is given as follows:

\begin{enumerate}
    \item \textbf{Parse Protein Structures}: Parse reference and generated protein structure files
    
    \item \textbf{Extract Top 7 Subsequences}: Identify the 7 longest subsequences (TM regions) of alpha carbons from the reference structure.
    
    \item \textbf{Find Corresponding Subsequences}:
    \begin{enumerate}
        \item For each reference subsequence, scan the generated structure to create sequences of alpha carbons with residue names and numbers.
        \item For each potential subsequence in the generated structure:
        \begin{itemize}
            \item Compare it with the reference subsequence.
            \item Count how many residues match in terms of residue names.
            \item Track the subsequence with the highest number of matching residues as the best match.
        \end{itemize}
    \end{enumerate}
    
    \item \textbf{Compute Distances}: Calculate the Euclidean distances between corresponding CA atoms of matching residues for each matched subsequence pair.
    
    \item \textbf{Calculate Average Deformation}: Determine the average deformation distance for each protein and store the results.
  
\end{enumerate}

This ensures that the correct residues are selected in the reference and generated structures for calculating the average deformation.

\section{4: t-SNE Visualization of GPCR Classes}

\begin{figure}[H]
    \centering
    \includegraphics[width = 16cm]{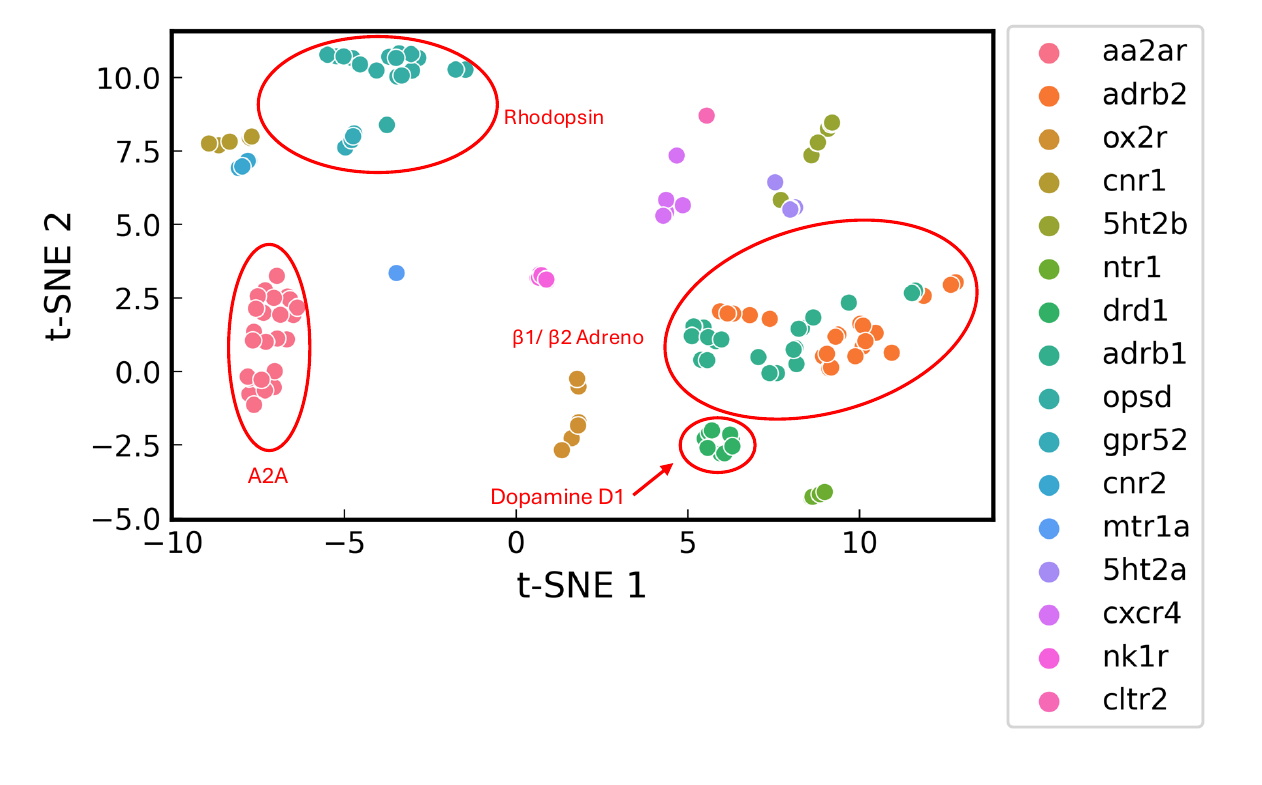}
    \caption{t-SNE visualization of GPCR-BERT embeddings. Clusters correspond to GPCR classes.}
    \label{fig:tsne_class}
\end{figure}

We refer to the t-SNE analysis conducted for GPCR-BERT, a language model for extracting higher-order interactions in GPCRs. By applying t-SNE to the embeddings from GPCR-BERT's final hidden state, we can visualize distinct clusters corresponding to different GPCR classes. Figure \ref{fig:tsne_class} shows the t-SNE results with clusters corresponding to different GPCR classes. Each point is color-coded according to its corresponding GPCR name. We should note that the dataset for this analysis (16 GPCRs) is significantly smaller than the dataset used for our study (75 GPCRs).

\clearpage
\bibliography{reference}